\documentclass[a4paper]{article}
\usepackage{color}
\usepackage{url}
\usepackage{multirow}

\usepackage{INTERSPEECH2019}

\title{Speaker Anonymization Using X-vector and Neural Waveform Models}
\name{Fuming Fang$^1$, Xin Wang$^1$, Junichi Yamagishi$^1$, Isao Echizen$^1$, \\ Massimiliano Todisco$^2$, Nicholas Evans$^2$, Jean-Fran\c{c}ois Bonastre$^3$}
\address{
  $^1$National Institute of Informatics, Tokyo, Japan\\
  $^2$EURECOM, France 
  $^3$University of Avignon, France}
\email{\{fang,wangxin,jyamagis,iechizen\}@nii.ac.jp, \{todisco,evans\}@eurecom.fr,
jean-francois.bonastre@univ-avignon.fr}

\begin{document}

\maketitle
\begin{abstract}
The social media revolution has produced a plethora of web services to which users can easily upload and share multimedia documents. Despite the popularity and convenience of such services, the sharing of such inherently personal data, including speech data, raises obvious security and privacy concerns. In particular, a user's speech data may be acquired and used with speech synthesis systems to produce high-quality speech utterances which reflect the same user's speaker identity. These utterances may then be used to attack speaker verification systems. One solution to mitigate these concerns involves the concealing of speaker identities before the sharing of speech data. For this purpose, we present a new approach to speaker anonymization. The idea is to extract linguistic and speaker identity features from an utterance and then to use these with neural acoustic and waveform models to synthesize anonymized speech. The original speaker identity, in the form of timbre, is suppressed and replaced with that of an anonymous pseudo identity. The approach exploits state-of-the-art x-vector speaker representations. These are used to derive anonymized pseudo speaker identities through the combination of multiple, random speaker x-vectors. Experimental results show that the proposed approach is effective in concealing speaker identities. It increases the equal error rate of a speaker verification system while maintaining high quality, anonymized speech.

\end{abstract}
\vspace{2mm}

\section{Introduction}
The development of the Internet has made it easy to share and acquire speech data. It is also easy to build speech synthesis systems from acquired data and then to use advanced speech synthesis techniques~\cite{lorenzo2018can} to generate new speech samples which reflect the voice of a specific speaker.
However, the abuse of these technological advancements has created new risks. Specially, generated utterances
could be used to attack an automatic speaker verification (ASV) system~\cite{wu2015asvspoof,kinnunen2017asvspoof,8630764}. 
It is also possible to search for information about a person on the Internet by using an ASV system~\cite{vestman2019sound}.
How to suppress
the identity information of a speaker and mitigate the risks is becoming an urgent problem.

The hiding of speaker identity, also referred to as {\it speaker anonymization} or {\it de-identification}, is a technology that modifies the original speech signal to make it sound like an anonymous
speaker's speech while maintaining the linguistic contents and speech quality. Speaker identity information typically includes timbre, pitch, speaking rate, and speaking style. Timbre, which can be represented by the spectrum, carries most of the speaker identity information and is widely used in ASV systems. We thus focused on modifying the timbre and developed a speaker anonymization method based on such modification.

The main idea of our proposed method is to separate speaker identity and the linguistic contents from the input speech and synthesize a new speech waveform after changing the speaker identity. As an initial trial, we focused on hiding speaker identity and maintaining speech quality by sacrificing a small part of linguistic contents. Specifically, we used a deep neural network (DNN)-based speaker-independent automatic speech recognition (ASR) system to capture linguistic information in the form of a phoneme posteriorgram (PPG)~\cite{sun2016phonetic}
and used a pre-trained x-vector~\cite{snyder2018x} system to encode the speaker identity. An anonymised pseudo speaker
is composed by combining the x-vectors corresponding to a set of different, arbitrary speakers.
Given the PPG and anonymized x-vector, our method uses neural acoustic and waveform models to generate an anonymized speech waveform. Experimental results show that speech anonymized using the proposed method is effective in concealing speaker identity, with anonymized speech being indistinguishable from the original speaker identity. Furthermore, anonymization protects speech quality, resulting in only negligible degradation.\footnote{A demonstration of audio samples is available at https://nii-yamagishilab.github.io/SpeakerAnonymization/}

The rest of this paper is organized as follows. Section~\ref{sec:related_works} briefly summarizes related work. Section~\ref{sec:proposed_method} gives the details of the proposed speaker anonymization method. Sections~\ref{sec:setup} and \ref{sec:res} describe the experimental setup and present the results. Finally, Section~\ref{sec:conclusion} summarizes the key points and mentions future work.

\section{Related work}
\label{sec:related_works}
First of all, speaker anonymization differs from speech anonymization~\cite{4284708,akagi2012privacy} in that the former suppresses speaker identity while the latter obscures linguistic content. In accordance with the manipulation objectives, speaker anonymization can be split into physical and logical anonymization. Physical anonymization aims to perturb speech in physical space by adding an external sound to the original waveform~\cite{hashimoto2016privacy} while logical anonymization modifies speaker identity on the recorded speech signal. Our proposed method falls into the latter category.

For logical anonymization, Jin et al.~\cite{jin2009speaker} presented a voice transformation (VT) system to change the speaker identity into another {\it special speaker}. Similarly, Bahmaninezhad et al.~\cite{Bahmaninezhad2018} utilized a convolutional neural network (CNN) as a VT function and averaged different transformation results as a means to anonymize speech. Magarienos et al.~\cite{magarinos2017reversible} and Pobar and Ip\v{s}i\'c~\cite{pobar2014online} improved the convenience of the VT-based method to enable the user to select an approximate transformation from a pool of pre-trained VT models for speaker anonymization. Our method differs from these methods because we train only one transformation function and generate an anonymised pseudo speaker identity through the combination of multiple, external speaker identities.

Justin et al.~\cite{justin2015speaker} performed speaker anonymization by first recognizing the diphones in the input speech using an ASR system and then synthesizing speech from the recognized diphone sequence. The synthesized speech differs from the original one in terms of speaker identity because the synthesizer is speaker-dependent and was trained using the data of a different speaker. This method is similar to our proposed method, but we use a speaker-independent speech synthesizer trained on the data for many speakers.  Our framework is therefore more flexible.

With a goal closely related to that of anonymization, Alegre et al.~\cite{alegre2014evasion} investigated so-called speaker evasion and obfuscation using 
voice conversion techniques. With the work aiming only to circumvent surveillance systems, it evaluated only how the approach could degrade ASV performance.  It did not consider degradations to speech quality. In contrast, the ideas presented in this paper are evaluated in terms of  
speaker identity anonymization, speech quality and linguistic content.

\section{Proposed speaker anonymization method}
\label{sec:proposed_method}
We assume that the information in a speech waveform can be disentangled and encoded using two groups of features. 
One group mainly encodes the speech content, e.g., the sequence of spoken words. The other group captures the acoustic features invariant to the speech content, e.g., the speaker identity. We further assume that we can anonymize a speech waveform by altering the features that encode the speaker identity only.

\begin{figure}[t]
\centering
\includegraphics[width=\columnwidth]{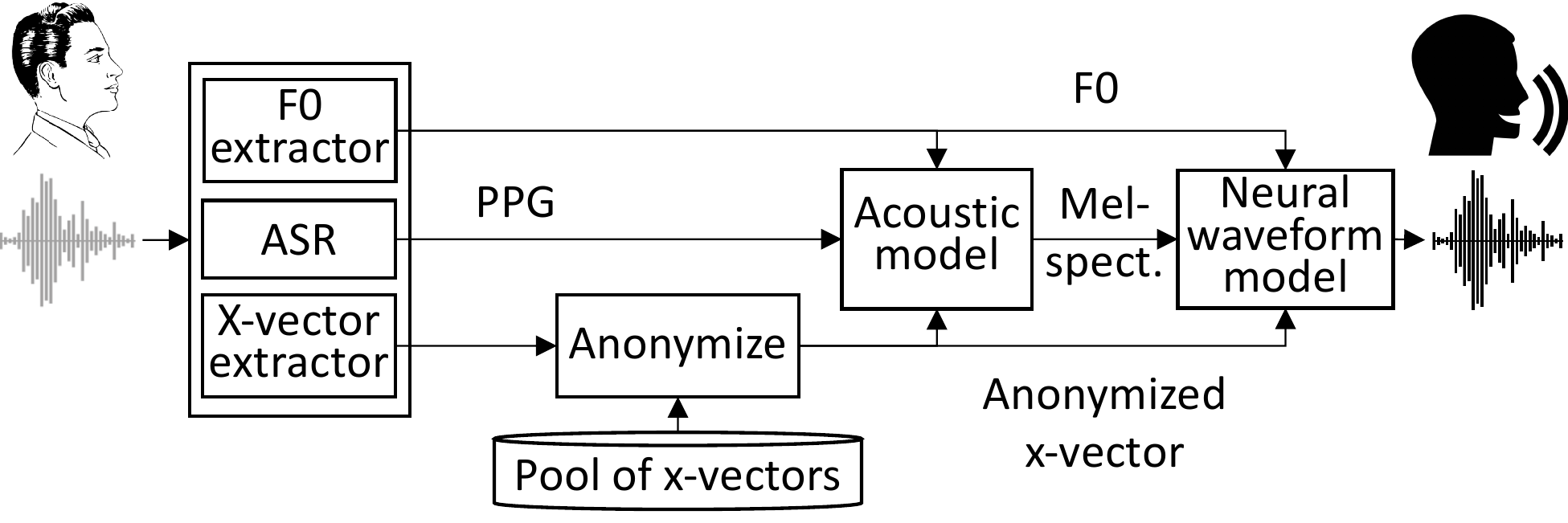}
\vspace{-7mm}
\caption{The proposed speaker anonymization system. PPG and Mel-spec. denote phoneme posteriorgram and Mel-spectrogram, respectively.}
\label{fig:system_overall}
\vspace{-5mm}
\end{figure}

Given these assumptions, we devised the speaker anonymization system illustrated in Figure~\ref{fig:system_overall}. This system first extracts an x-vector, a PPG, and the fundamental frequency (F0) from the input waveform. It then anonymizes the x-vector on the basis of information gleaned from the x-vectors of external speakers. Finally, it uses an acoustic model and a neural waveform model to synthesize the speech waveform from the anonymized x-vector and the original PPG and F0.

\subsection{Feature extraction}
X-vector representations are used as features to encode speaker identity because they have been shown to be effective in speaker verification and recognition systems~\cite{snyder2018x}. The x-vector extractor is a DNN consisting of seven fully connected layers, a stats pooling layer, and a softmax output layer, as listed in Table~\ref{tbl:xvector}. This DNN takes 24-dimensional filter banks as input, and the first three layers splice several frames of the previous layer's output as its input. As a result, the third layer can extract one feature vector that covers 15 input frames of the filter bank features. To deal with the varied length of the input speech, a stats pooling layer is used after the fifth layer to calculate the mean and variance of the overall output of the fifth layer. The softmax layer predicts the probability that the input speech is from each of the $N$ speakers in the training data. Once the network is trained on a database containing a large number of speakers, the outputs of the higher layers can be used to represent the speaker space, and the trained model can be used to extract the speaker identity for a new input speech signal. We followed the standard approach for x-vector extraction, i.e., by using the affine component of the sixth layer~\cite{snyder2018x}. Finally, the utterance-level x-vectors of a speaker are averaged as the speaker-level x-vector.

\begin{table}[tb]
\caption{Network architecture of x-vector system. $T$ denotes speech length, $t$ indicates $t$-th frame, $\{\cdot\}$ is a set of frame indices, and $N$ is the number of speakers in training data.}
\vspace{-6mm}
\begin{center}
\begin{tabular}{|c|c|c|c|}
\hline
Layer & Layer context & Context & Input $\times$ output \\
\hline
1 & $[t-2, t+2]$ & 5 & 120 $\times$ 512 \\
2 & $\{t-2, t, t+2\}$ & 9 & 1536 $\times$ 512 \\
3 & $\{t-3, t, t+3\}$ & 15 & 1536 $\times$ 512 \\ 
4 & $\{t\}$ & 15 & 512 $\times$ 512 \\
5 & $\{t\}$ & 15 & 512 $\times$ 1500 \\
stats & $[0,T)$ & $T$ & $1500T \times 3000$ \\
6 & $\{0\}$ & $T$ & $3000 \times 512$ \\
7 & $\{0\}$ & $T$ & $512 \times 512$ \\
softmax & $\{0\}$ & $T$ & $512 \times N$ \\
\hline
\end{tabular}
\end{center}
\label{tbl:xvector}
\vspace{-7mm}
\end{table}

The PPG is used as a feature representation for encoding the linguistic content. The PPG is a sequence of vectors that contain the posterior probability of every phoneme class at the corresponding time step (or speech frame). In other words, each PPG vector is a soft label indicating the likelihood of each possible phoneme being uttered in the speech frame. For this work, we use a DNN-based ASR system~\cite{kaldi_timit} to extract the PPG. The network is a stack of six sigmoid layers and a softmax output layer. Each input feature vector is the concatenation of 11 frames of 40-dimensional acoustic feature vectors. The hidden layers have a layer size of 1024, while the layer size of the softmax layer is 1944, i.e., the number of tied tri-phone hidden Markov model states. Because this ASR system is trained on a database containing a large number of speakers, the learned DNN can be treated as a speaker-independent model. We assume that the extracted PPG or output of the top hidden layers mainly encodes the speech content rather than the speaker identity. In addition to the softmax layer, we also consider the output of the 6th sigmoid layer as a PPG.

The proposed system extracts F0 from the input speech waveform using an F0 extractor ensemble~\cite{juvelaniibc}. Although the F0 may encode speaker identity information such as gender and age, 
it also contains plenty of context-related information such as the pitch accents and intonation that carry the semantic message~\cite{gussenhoven2004phonology}. Therefore, in order to preserve the context information, the proposed system does not modify F0.
Furthermore, ASV systems normally use short-time spectral features rather than F0 to verify speaker identity.

Note that the F0 is extracted with a frame shift of 5 ms. Because the ASR system produces a PPG vector every 10 ms, the extracted PPG vector is replicated twice to match the F0 frame rate. The x-vector is copied to every frame.

\subsection{Anonymization}
\begin{figure}[tb]
\begin{tabular}{cc}
\begin{minipage}{0.5\hsize}
\begin{center}
\includegraphics[width=0.9\columnwidth]{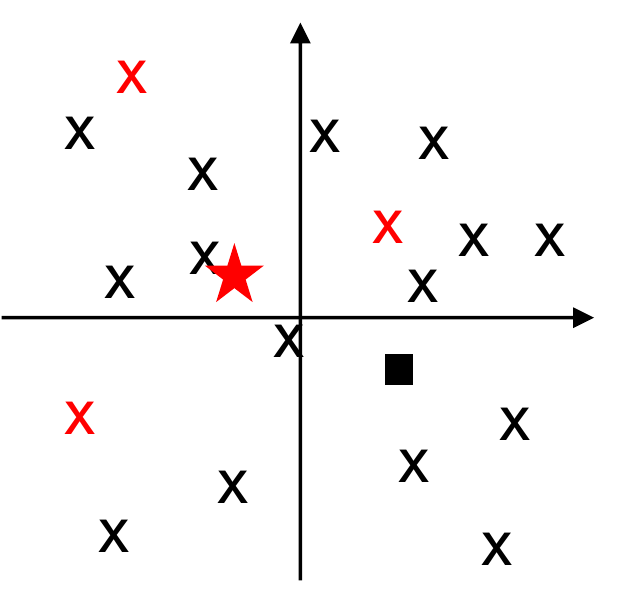}\\
(a) Random selection.
\end{center}
\end{minipage}
\begin{minipage}{0.5\hsize}
\begin{center}
\includegraphics[width=0.9\columnwidth]{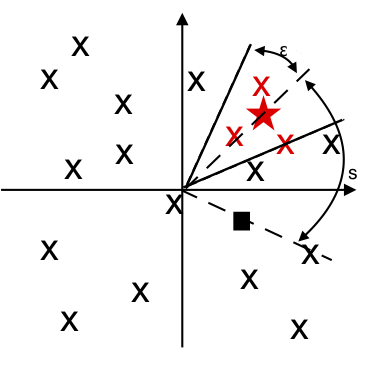}\\
(b) Range selection.
\end{center}
\end{minipage}
\end{tabular}
\caption{Schematic representations of speaker anonymization based on two x-vector selection methods. Square markers indicate target speaker to be anonymized, red cross markers indicate selected x-vectors, and red star markers represent average of selected x-vectors.}
\label{fig:anonymization_methods}
\vspace{-5mm}
\end{figure}

Many methods can be used to generate a new speaker identity on the basis of the x-vector. We devised two simple anonymization methods for modifying the x-vector of the input speech waveform. One is to use the mean x-vector of a set of randomly selected x-vectors from an x-vector pool, which produces a different anonymized psuedo speaker each time. The other is to compose an x-vector for which the similarity score to the original x-vector is $s$. The latter method enables the distance between the anonymized speaker and the original input speaker to be flexibly controlled. The composed x-vector can be approximately calculated by averaging a set of candidate x-vectors for which the similarity to the original speaker is in the range $[s-\epsilon, s+\epsilon]$, where $\epsilon > 0$ is a hyper-parameter used to control the width of the range. We use cosine distance $cos({\bf x}_1, {\bf x}_2)$ to represent the similarity between two x-vectors ${\bf x}_1$ and ${\bf x}_2$.
These two anonymization methods are illustrated in Figure~\ref{fig:anonymization_methods}.

\subsection{Waveform generation}
The proposed system uses two modules to generate the speech waveform: an acoustic model that generates a Mel-spectrogram given the three input features (PPG, F0, and anonymized x-vector), and a neural source-filter (NSF) waveform model~\cite{wang2018neural} that produces a speech waveform given the F0, anonymized x-vector, and generated Mel-spectrogram.

The acoustic model is an autoregressive (AR) one that generates one frame of a Mel-spectrogram given the data generated in the previous frames. Similar to the acoustic model in our previous work~\cite{lorenzo2018can}, this AR acoustic model has two feedforward layers and a bi-directional long short-term memory (LSTM) recurrent layer near the input side. It uses a unidirectional LSTM layer that takes the output of the previous LSTM layer and the Mel-spectrogram generated in the previous frame as its input. Finally, a linear output layer is used to produce the Mel-spectrogram of the current time step. During training, the natural Mel-spectrogram is fed back into the unidirectional LSTM layer (i.e., teacher forcing). The dimension of each Mel-spectrogram vector is 80.

\begin{figure}[t]
\centering
\includegraphics[width=\columnwidth]{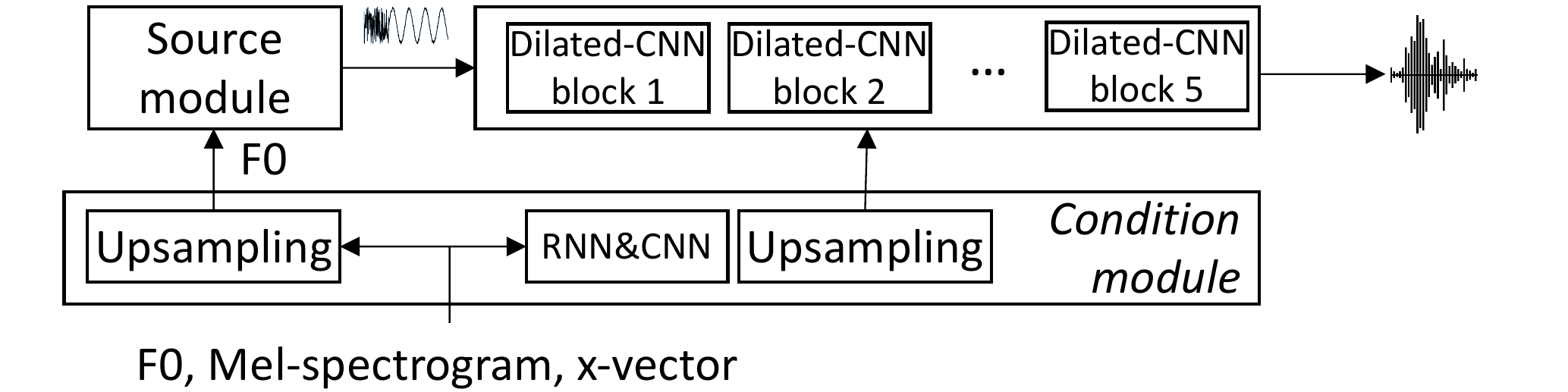}
\vspace{-6mm}
\caption{System diagram of neural source-filter-based waveform model. }
\label{fig:system_nsf}
\end{figure}

The NSF model mainly contains three modules, as shown in Figure~\ref{fig:system_nsf}. The condition module processes the input features and upsamples them to the waveform sampling rate. The source module generates a sine-waveform excitation signal given the upsampled F0, and the filter module converts the excitation into a waveform using five dilated-convolution (dilated-CNN) blocks. This model is trained by minimizing the short-time spectral amplitude distances between the generated and natural waveforms. We used our previous specifications \cite{wang2018neural} to configure it. Specifically, each of the five dilated-CNN blocks contains ten dilated-CNN layers, where the dilation size of the $k$-th layer is $2^{k-1}$. 

While the x-vector extractor and ASR modules can be pre-trained using external sources, the acoustic model and the NSF model must be trained using data from potential users of the anonymization system.

\section{Experimental setup}
\label{sec:setup}
\subsection{Evaluation methods}
We evaluated the performance of the proposed system using an ASV system (ASV$_\text{eval}$) and an ASR system (ASR$_\text{eval}$).  The latter is different to the ASR system used for PPG extraction (ASR$_{\text{ppg}}$). Suppose a speaker has been enrolled in the ASV$_\text{eval}$ system. Ideally, the anonymized speech of the speaker will be `falsely' rejected by the ASV$_\text{eval}$ system. By calculating how the rejection rates rise over multiple speakers, we evaluated how well the proposed system anonymized the speech waveforms. We used the ASR$_\text{eval}$ system to recognize the word sequences from the original and anonymized speech. It is assumed that the smaller the difference between the word sequences, the better the preservation of the linguistic contents.

We used the equal error rate (EER) to measure anonymization performance and the word error rate (WER) to investigate how well the content was preserved in the anonymized speech. We used the mean opinion score (MOS) as the metric for the quality of the anonymized speech. It was calculated using the P.563 algorithm~\cite{rec2004p} and evaluated by human listeners.

\begin{table}
\footnotesize
\begin{center}
\caption{Training recipes for modules in proposed system and other systems for performance evaluation.}
\vspace{-3mm}
\begin{tabular}{|p{11mm}|p{13mm}|p{44mm}|}
\hline
\multicolumn{2}{|c|}{Module} & Description \\
\hline\hline
\multirow{7}{10mm}{Proposed system} &  \multirow{2}{10mm}{ASR$_{\text{ppg}}$} & trained on TIMIT database using Kaldi TIMIT recipe~\cite{kaldi_timit} \\
\cline{2-3}

			   & x-vector extractor  & trained on VoxCeleb database using Kaldi VoxCeleb recipe \cite{snyder2018x} \\
			   \cline{2-3}
			   & x-vector pool & extracted from VoxCeleb database using x-vector extractor (7,325 speakers) \\
			   \cline{2-3}
            &  Acoustic model\&NSF  & trained on VCTK corpus using recipe in Table~\ref{table:data_recipe} \\
			   
\hline
\hline
\multirow{3}{12mm}{Evaluation modules} & ASR$_{\text{eval}}$ & pre-trained DeepSpeech model \cite{Hannun2014DeepSS} \\ \cline{2-3}
& \multirow{2}{10mm}{ASV$_{\text{eval}}$} & trained on VoxCeleb database using VoxCeleb ASV recipe \cite{nagrani2017voxceleb} and adapted to VCTK corpus  \\
\hline
\end{tabular}
\label{table:modules_recipe}
\end{center}
\vspace{-6mm}
\end{table}

\begin{table}[t]
\footnotesize
\begin{center}
\caption{Detailed usage of VCTK corpus in NSF and ASV$_\text{eval}$ modules. Spk denotes speaker ID, and utt. denotes utterance ID, which is shared across speakers. Tar. and nontar. denote target and non-target test data for ASV evaluation.
}
\vspace{-2mm}
\begin{tabular}{| p{11mm} | l | l | p{20mm} |}
\hline
    &    Data set & No. of speakers         &  No. of utterances \\
\hline
\hline
AM\&NSF  & train/dev  & 68 (spk.01 - 68)  & 13600 (utt.026 - 225) \\
\hline
\hline
\multirow{4}{*}{ASV$_\text{eval}$} & adaptation & 20 (spk.69 - 89)         & 2580 (utt.226 -  ) \\ \cline{2-4}
    & enrollment & 11 (spk.01 - )       &  129 (utt.001 - )  \\ \cline{2-4}
& test (tar.)    & 11 (spk.01 - )      & 1528 (utt.226 - ) \\ \cline{2-4}
    & test (nontar.) & 19 (spk.49 - 67)   & 1985 (utt.226 - )  \\    
\hline
\end{tabular}
\label{table:data_recipe}
\end{center}
\vspace{-6mm}
\end{table}

\subsection{Data recipes}
As listed in Table~\ref{table:modules_recipe}, the proposed system contained modules trained using different data corpora. The ASR$_{\text{ppg}}$ module is a DNN-based phoneme recognizer system trained on the TIMIT database \cite{garofolo1993timit}. The x-vector extractor was configured and trained on the VoxCeleb dataset using the Kaldi VoxCeleb training recipe \cite{nagrani2017voxceleb}. With the x-vector extractor, the pool of x-vectors for anonymization was extracted from the VoxCeleb dataset, which resulted in 7,325 x-vectors. The acoustic model and the NSF module were trained on the VCTK corpus using the recipes in Table~\ref{table:data_recipe}. The training and development data included 13,600 utterances from 68 speakers, each of whom contributed 200 utterances. The ratio of training and development utterances was around $4:1$, and the numbers of male and female speakers were 30 and 38, respectively.

The performance of the proposed system was evaluated on a test data set
(11 American speakers randomly selected from the 68 speakers), and the remaining 19 speakers were non-target speakers (10 female and 9 male). The number of utterances for the target and non-target speakers are listed in Table~\ref{table:data_recipe}.

As the evaluator, the x-vector-PLDA-based ASV$_\text{eval}$ ~\cite{prince2007probabilistic} module was trained on the VoxCeleb dataset and then adapted to the VCTK domain using 2580 utterances from 20 unused speakers in the VCTK corpus. These 20 speakers differed from the target and non-target speakers. After training and adaptation, each of the 11 target speakers enrolled in ASV$_\text{eval}$ using at most 19 enrollment utterances. There was no overlap among the enrollment, adaptation, training, and test datasets, as shown in Table~\ref{table:modules_recipe}. The EERs were calculated on the basis of a gender-dependent verification system. Finally, the ASR$_\text{eval}$ module was DeepSpeech\footnote{https://github.com/mozilla/DeepSpeech}~\cite{Hannun2014DeepSS} pre-trained on external data.

Note that, since the ASR$_{\text{ppg}}$ and DeepSpeech models were trained on American English, if they had been used directly for VCTK-based evaluation, the non-American speakers would have had large WERs. Therefore, we used only American speakers in the VCTK corpus as target speakers to be anonymized (especially for the ``test (tar.)'' dataset listed in Table~\ref{table:data_recipe}).

\subsection{Subjective evaluation setup}
We carried out a subjective evaluation test to investigate the quality of the anonymized speech and the similarity between the original natural speech and the anonymized speech. The range selection method was used for the anonymization. We used four levels of dissimilarity for each PPG condition (i.e., the 6th sigmoid and softmax layers): 0.0, 0.2, 0.4, 0.6, where 0.0 means copy synthesis. The evaluation was performed by 296 listeners who each rated up to 380 utterances randomly selected from the 1,528 test utterances. A combined total of 26,978 data points were obtained for the quality and similarity evaluations, which is roughly equivalent to each utterance being evaluated twice, respectively. 

\section{Experimental results}
\label{sec:res}
\subsection{Anonymization using nearest speakers}
Anonymized speech was obtained by averaging the nearest $M$ speakers' x-vectors in the pool, where $M=100$, 200, and 300. Because the `dissimilar' non-target speakers far from the target speaker might be assigned low scores, resulting in low EERs, we used the $K$ nearest non-target speakers for EER calculation, where $K=3$, 6, and all (9 male and 10 female speakers).

As shown in Table~\ref{tbl:eer}, the EERs before anonymization are low for all conditions (around 2.04\% to 2.52\%). After anonymization, they are much higher. The highest rates are achieved for a value of $M=300$ when the softmax layer output was used as the PPG. This indicates that the proposed method can effectively suppress speaker identity. Comparing the two PPG conditions, the softmax layer resulted in higher EERs. This might be because the PPG from the ASR's softmax layer contains less speaker identity information while that from the hidden layers contains more speaker-dependent features that may have been used by the acoustic model.

\begin{table}
    \begin{center}
    \caption{EERs (\%) for anonymized speech and nearest $K$ non-target speakers. Anonymized speaker was composed using $M$ nearest speakers in x-vector pool. ``$-$'' means no anonymization performed, ``6th sigmoid'' means 6th sigmoid hidden layer of ASR$_{\text{ppg}}$, and ``softmax'' is softmax output layer of ASR$_{\text{ppg}}$.}
    \vspace{-3mm}
    \begin{tabular}{|c|c|c|c|}
    \hline
    Anonymization using $M$ & \multicolumn{3}{c|}{$K$} \\ \cline{2-4}
    nearest speakers in pool & 3 & 6 & all \\
    \hline\hline
    $-$ & 2.52 & 2.04 & 2.04 \\
    \hline
    \hline
    \multicolumn{4}{|c|}{PPG: 6th sigmoid} \\
    \hline
100 & 11.69 & 10.11 & 9.56 \\
200 & 11.89 & 10.38 & 9.80 \\
300 & 12.23 & 10.73 & 10.11 \\
    \hline
    \hline
    \multicolumn{4}{|c|}{PPG: softmax} \\
    \hline
100 & 23.00 & 20.77 & 19.74 \\
200 & 24.81 & 21.92 & 21.66 \\
300 & 25.05 & 22.89 & 22.60 \\
    \hline
    \end{tabular}
    \label{tbl:eer}
    \end{center}
    \vspace{-3mm}
\end{table}

\subsection{Anonymization using random selection method}
\begin{table}
    \begin{center}
    \caption{EERs (\%) for anonymized speakers (using $M$ randomly selected speakers from pool) and non-target speakers.}
    \vspace{-3mm}
    \begin{tabular}{|c|c|c|c|c|}
    \hline
    Anonymization with $M$ & \multirow{2}{*}{10} & \multirow{2}{*}{50} & \multirow{2}{*}{100} & \multirow{2}{*}{200} \\
    random speakers in pool &  &  &  & \\
    \hline
    $-$ & \multicolumn{4}{c|}{2.11} \\
    \hline
    PPG: 6th sigmoid & 32.17 & 22.66  &   23.20 &    23.89 \\
    \hline
    PPG: softmax & 34.28 & 30.35  &   29.65 & 29.46 \\
    \hline
    \end{tabular}
    \label{tbl:random_eer}
    \end{center}
    \vspace{-6mm}
\end{table}
\begin{table}
    \vspace{-3mm}
    \begin{center}
    \caption{Dissimilarity score range between original speaker and anonymized speaker composed using $M$ randomly selected speakers.}
    \vspace{-3mm}
    \begin{tabular}{|c|c|c|c|c|}
    \hline
    $M$ & 10 & 50 & 100 & 200 \\
    \hline
    6th & \multirow{2}{*}{0.17-0.53} & \multirow{2}{*}{0.20-0.43} & \multirow{2}{*}{0.19-0.44} & \multirow{2}{*}{0.21-0.41} \\
    sigmoid & &&& \\
    \hline
    softmax & 0.17-0.60 & 0.26-0.51 & 0.24-0.51 & 0.26-0.48 \\
    \hline
    \end{tabular}
    \label{tbl:score_range}
    \end{center}
    \vspace{-6mm}
\end{table}

We randomly selected $M$ speakers from the pool for anonymization, where $M=10$, 50, 100, and 200. To reduce the bias caused by random selection, we repeated the experiment five times for each $M$, and the average EER was used as the final result. The EERs for the anonymized speakers and the non-target speakers are shown in Table~\ref{tbl:random_eer}. Compared with the original speaker, the anonymized speakers achieved much higher EERs. This means that random selection is a simple and effective method for anonymization.
Note that EERs for the original speakers in Table~\ref{tbl:eer} (2.04\%) and Table~\ref{tbl:random_eer} (2.11\%) were different. This was because of different protocols used.

Table~\ref{tbl:score_range} shows the dissimilarity score ranges for the anonymized speaker and the original speaker. The scores were calculated from their x-vectors, and the range for each $M$ was calculated from the scores for the five repetitions. The dissimilarity score is defined as $1-cos({\bf x}_1, {\bf x}_2)$. As shown by these results, the random selection produced various anonymized speakers for the original target speakers.

\subsection{Anonymization with range selection method}
\begin{figure}[t]
\centering
\includegraphics[width=0.8\columnwidth]{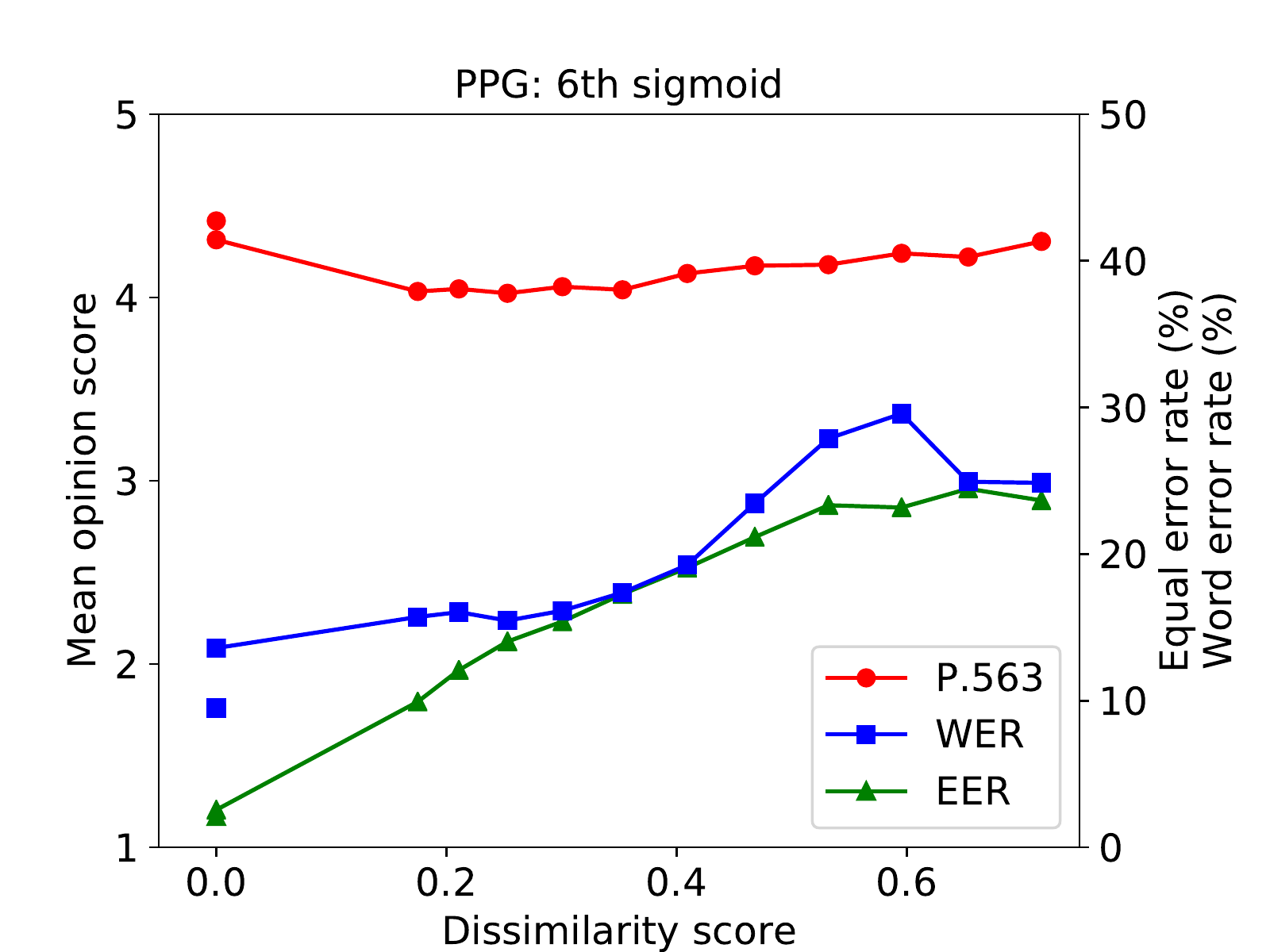}
\includegraphics[width=0.8\columnwidth]{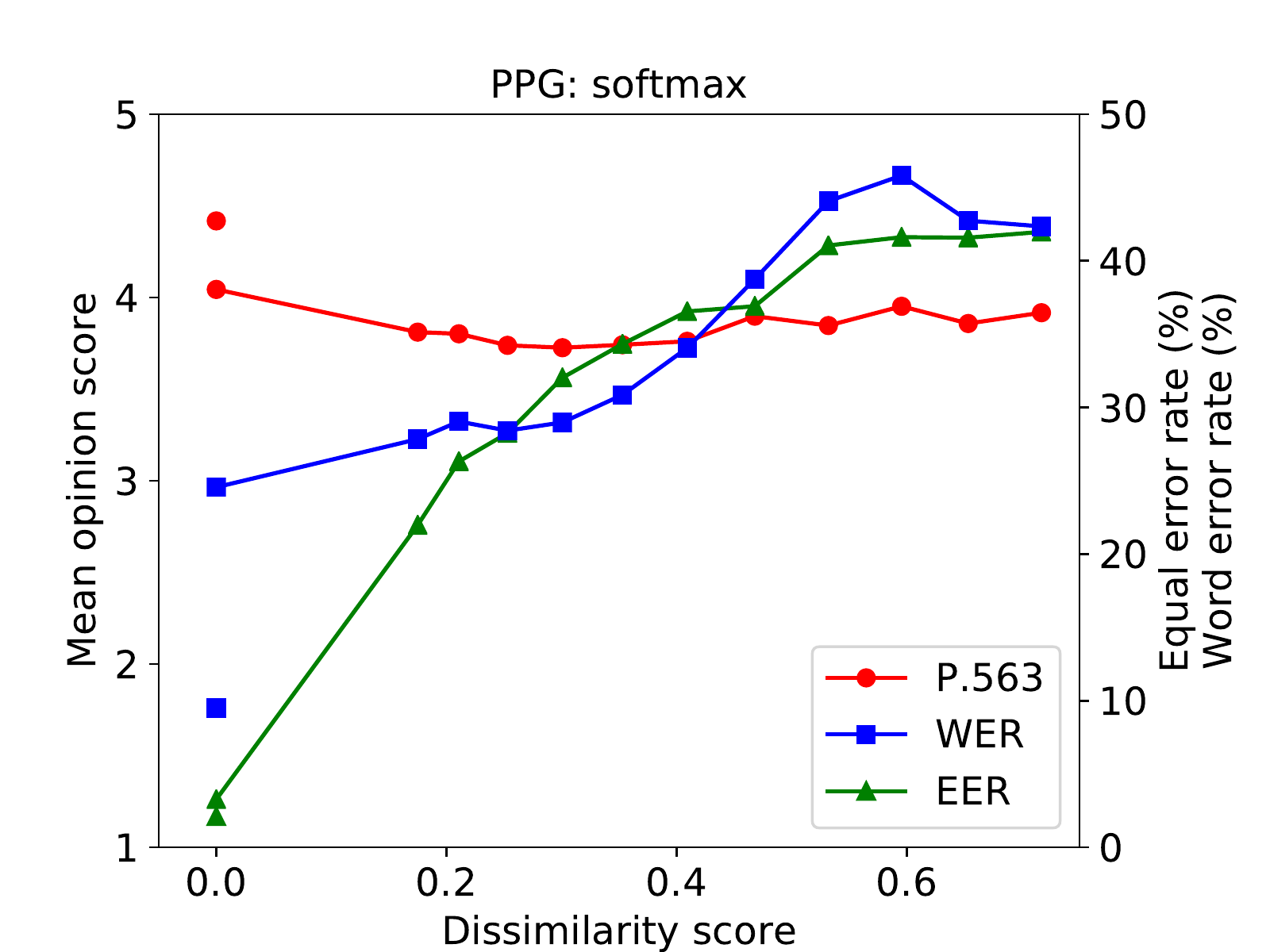}
\vspace{-2mm}
\caption{EER, WER, and MOS obtained by anonymization with different distances. 
``0.0'' means without anonymization or copy-synthesis.}
\label{fig:ex_distance}
\vspace{-5mm}
\end{figure}

Figure~\ref{fig:ex_distance} shows the EERs, WERs, and MOS values obtained by anonymization with different distances between the anonymized speaker and the original input speaker. We use the dissimilarity score to represent the distance. As the dissimilarity score increases, the EER increases greatly while the MOS is relatively stable. This indicates that the proposed method can generate anonymized speakers while maintaining speech quality.

However, there was a large gap between the natural speech and the anonymized speech in terms of the WER. The WER of the natural speech is 9.49\% while those of the anonymized speech are between 10\% and 30\% when using the PPG from the 6th layer and between 25\% and 45\% when using the PPG from the softmax layer. 

One reason for the higher WERs may be that the ASR${\text{ppg}}$ cannot perfectly recognize the phoneme sequence from the input speech signal and may extract inaccurate PPGs. Such inaccurate PPGs could deliver incorrect linguistic contents and thus affect the acoustic models both in training and generation. This suggests that we need to train an ASR$_{\text{ppg}}$ using a larger amount of data that include many accents or to use unsupervised representation learning frameworks to obtain phone-equivalent information without using phone knowledge. This is the next step in our work.

In addition, we can see that the WERs were higher when the dissimilarity score was higher than 0.4. This indicates that when x-vectors are averaged over very different unseen speakers, our proposed system is unable to correctly recover the original linguistic contents of the input speech. This means that, while anonymization is largely successful, the speaker identity and linguistic contents are not perfectly factorized in the current framework. Accordingly, we need to devise a framework that can better disentangle their representations. 

\subsection{Subjective test}
\begin{figure}[t]
\centering
\includegraphics[width=0.8\columnwidth]{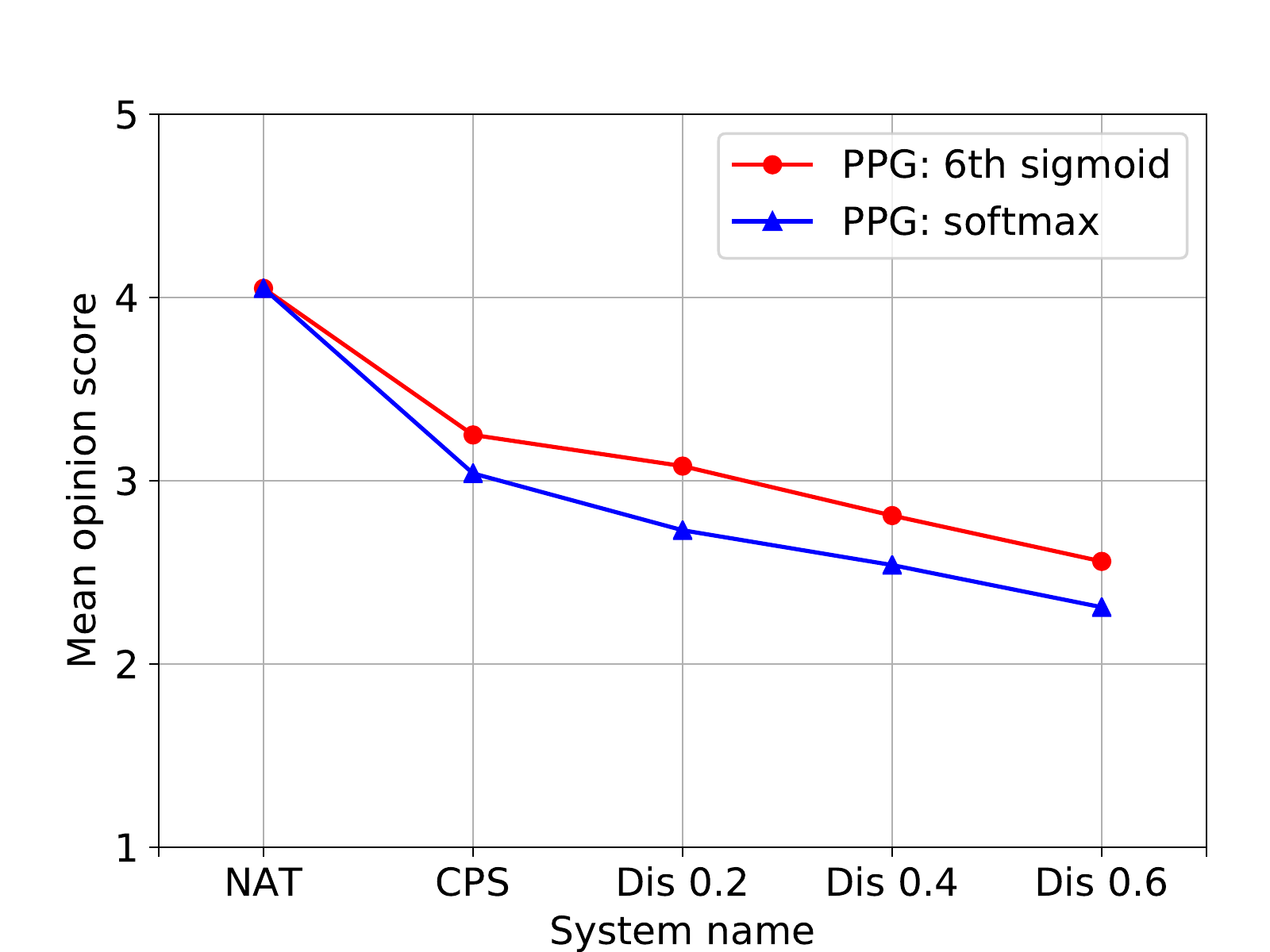}
\vspace{-2mm}
\caption{Speech quality in terms of human listener MOS. NAT: original natural speech, CPS: copy synthesis, Dis: dissimilarity score.}
\label{fig:subjective_mos}
\vspace{-3mm}
\end{figure}
\begin{figure}[t]
\centering
\includegraphics[width=0.8\columnwidth]{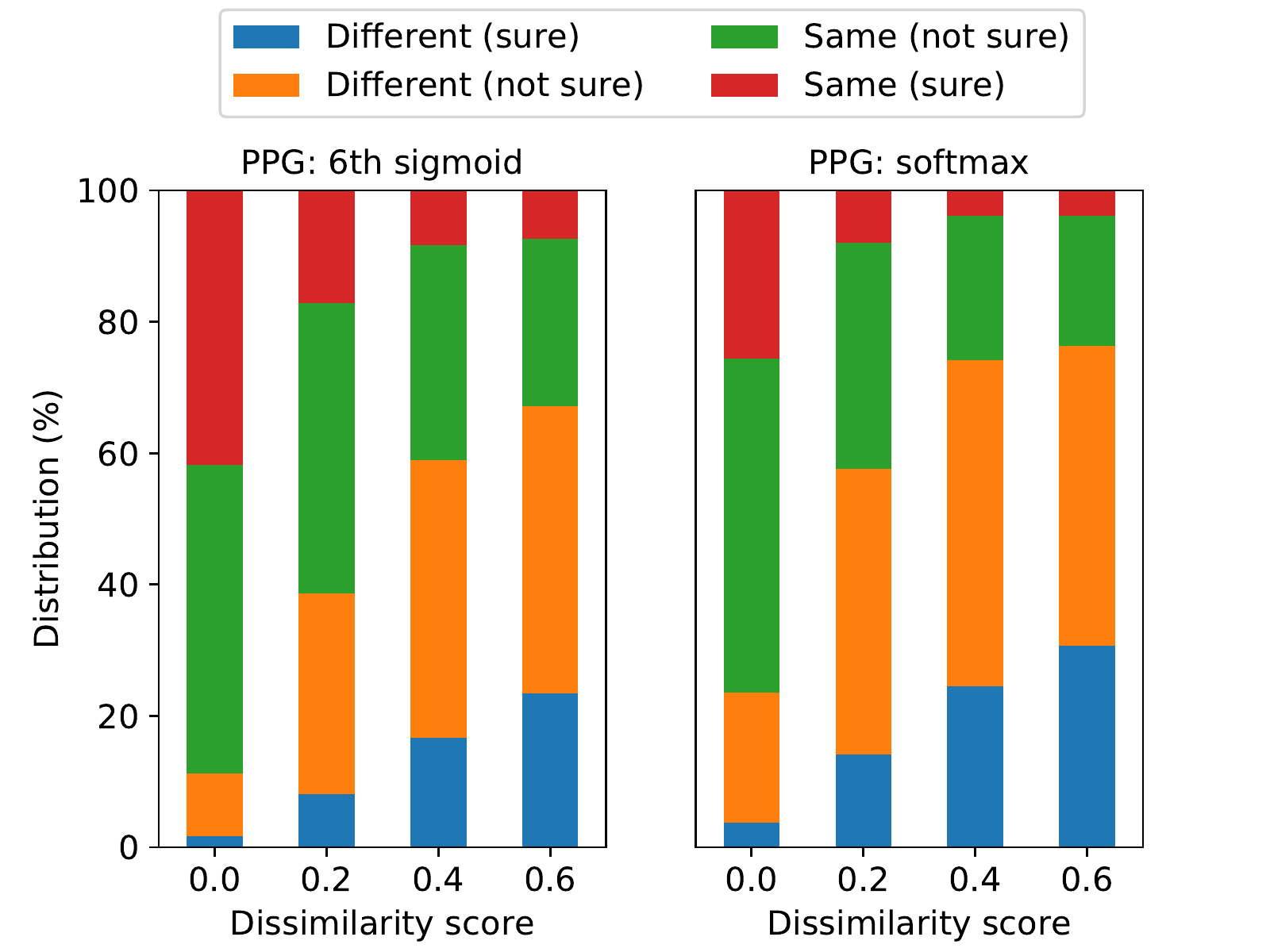}
\vspace{-2mm}
\caption{Similarity between the original speaker and the anonymized speaker broken down by evaluator judgment. Dissimilarity score of 0.0 means copy synthesis.}
\label{fig:subjective_similarity}
\vspace{-2mm}
\end{figure}

As shown in Figure~\ref{fig:subjective_mos}, the original natural speech had an average MOS of 4.05 whereas the copy synthesized and anonymized speech scores were between 2.31 and 3.25. The scores slightly decreased as the dissimilarity score increased. The reason for the copy synthesized and anonymized speech scores being worse than the natural speech score might be that linguistic errors and the untangled speaker identity blurred the synthesized speech. 

Figure~\ref{fig:subjective_similarity} breaks down the similarity between the original speaker and the anonymized speaker by evaluator judgment. As the dissimilarity score increased, the judgment as different increased. This indicates that the proposed method can anonymize speech sufficiently well to hide the speaker’s identity. 

As we can see from Figures~\ref{fig:subjective_mos} and \ref{fig:subjective_similarity}, higher quality was achieved when the PPG was extracted from the 6th sigmoid layer while more speaker-distinguishable speech was produced when it was extracted from the softmax layer. These results are similar to the objective results above. 

\section{Conclusion and future work}
\label{sec:conclusion}
Our proposed speaker anonymization method is based on x-vectors and neural acoustic and waveform models. The essential idea is to first extract speaker identity and linguistic information from the audio signal and then synthesize anonymized speech from the linguistic information and an anonymised pseudo speaker identity. The anonymised pseudo speaker is composed by averaging multiple speakers' x-vectors, and the PPG is used as a representation of the linguistic information. Experimental results showed that the use of anonymized speech greatly increased the EER of an automatic speaker verification system while the quality of the anonymized speech was only slightly degraded. However, the anonymized speech had higher WERs. This was probably caused by a phoneme recognition error from the ASR system used for PPG extraction. 

Since the proposed method heavily depends on the accuracy of the ASR system for PPG extraction, we plan to develop an unsupervised method to disentangle the linguistic contents and speaker identity and thus avoid the error introduced by the ASR system.
Since no common databases, protocols, and metrics are available, it is difficult to compare different solutions and we did not have a baseline system in this paper. Therefore, we plan to provide a common database, protocol, and metric for development of speaker anonymization. 

\section{Acknowledgements}
This work was partially supported by a JST CREST Grant (JPMJCR18A6, VoicePersonae project), Japan, and by MEXT KAKENHI Grants (16H06302, 17H04687, 18H04120, 18H04112, 18KT0051), Japan. The numerical calculations were carried out on the TSUBAME 3.0 supercomputer at the Tokyo Institute of Technology.

The authors would like to thank Mr.\ Ville Vestman of the University of Eastern Finland for kindly providing a PLDA ASV system.

\bibliographystyle{IEEEtran}

\bibliography{refs}

\begin{thebibliography}{10}
\providecommand{\url}[1]{#1}
\csname url@samestyle\endcsname
\providecommand{\newblock}{\relax}
\providecommand{\bibinfo}[2]{#2}
\providecommand{\BIBentrySTDinterwordspacing}{\spaceskip=0pt\relax}
\providecommand{\BIBentryALTinterwordstretchfactor}{4}
\providecommand{\BIBentryALTinterwordspacing}{\spaceskip=\fontdimen2\font plus
\BIBentryALTinterwordstretchfactor\fontdimen3\font minus
  \fontdimen4\font\relax}
\providecommand{\BIBforeignlanguage}[2]{{%
\expandafter\ifx\csname l@#1\endcsname\relax
\typeout{** WARNING: IEEEtran.bst: No hyphenation pattern has been}%
\typeout{** loaded for the language `#1'. Using the pattern for}%
\typeout{** the default language instead.}%
\else
\language=\csname l@#1\endcsname
\fi
#2}}
\providecommand{\BIBdecl}{\relax}
\BIBdecl

\bibitem{lorenzo2018can}
J.~Lorenzo-Trueba, F.~Fang, X.~Wang, I.~Echizen, J.~Yamagishi, and T.~Kinnunen,
  ``{Can we steal your vocal identity from the Internet?: Initial investigation
  of cloning Obama's voice using GAN, WaveNet and low-quality found data},'' in
  \emph{Speaker Odyssey, The Speaker and Language Recognition Workshop}, 2018.

\bibitem{wu2015asvspoof}
Z.~Wu, T.~Kinnunen, N.~Evans, J.~Yamagishi, C.~Hanil{\c{c}}i, M.~Sahidullah,
  and A.~Sizov, ``{ASVspoof 2015: the first automatic speaker verification
  spoofing and countermeasures challenge},'' in \emph{Sixteenth Annual
  Conference of the International Speech Communication Association}, 2015.

\bibitem{kinnunen2017asvspoof}
T.~Kinnunen, M.~Sahidullah, H.~Delgado, M.~Todisco, N.~Evans, J.~Yamagishi, and
  K.~A. Lee, ``{The ASVspoof 2017 challenge: Assessing the limits of replay
  spoofing attack detection},'' \emph{ISCA (the International Speech
  Communication Association)}, 2017.

\bibitem{8630764}
F.~{Fang}, J.~{Yamagishi}, I.~{Echizen}, M.~{Sahidullah}, and T.~{Kinnunen},
  ``Transforming acoustic characteristics to deceive playback spoofing
  countermeasures of speaker verification systems,'' in \emph{2018 IEEE
  International Workshop on Information Forensics and Security (WIFS)}, Dec
  2018, pp. 1--9.

\bibitem{vestman2019sound}
V.~Vestman, B.~Soomro, A.~Kanervisto, V.~Hautam{\"a}ki, and T.~Kinnunen, ``Who
  do i sound like? showcasing speaker recognition technology by youtube voice
  search,'' in \emph{ICASSP 2019-2019 IEEE International Conference on
  Acoustics, Speech and Signal Processing (ICASSP)}.\hskip 1em plus 0.5em minus
  0.4em\relax IEEE, 2019, pp. 5781--5785.

\bibitem{sun2016phonetic}
L.~Sun, K.~Li, H.~Wang, S.~Kang, and H.~Meng, ``Phonetic posteriorgrams for
  many-to-one voice conversion without parallel data training,'' in \emph{2016
  IEEE International Conference on Multimedia and Expo (ICME)}.\hskip 1em plus
  0.5em minus 0.4em\relax IEEE, 2016, pp. 1--6.

\bibitem{snyder2018x}
D.~Snyder, D.~Garcia-Romero, G.~Sell, D.~Povey, and S.~Khudanpur, ``{X-vectors:
  Robust DNN embeddings for speaker recognition},'' in \emph{2018 IEEE
  International Conference on Acoustics, Speech and Signal Processing
  (ICASSP)}.\hskip 1em plus 0.5em minus 0.4em\relax IEEE, 2018, pp. 5329--5333.

\bibitem{4284708}
J.~{Chen}, D.~T. {Huy}, K.~{Phua}, J.~{Biswas}, and M.~{Jayachandran}, ``Using
  keyword spotting and replacement for speech anonymization,'' in \emph{2007
  IEEE International Conference on Multimedia and Expo}, July 2007, pp.
  548--551.

\bibitem{akagi2012privacy}
M.~Akagi and Y.~Irie, ``Privacy protection for speech based on concepts of
  auditory scene analysis,'' \emph{Institute of Noise Control Engineering of
  the USA}, 2012.

\bibitem{hashimoto2016privacy}
K.~Hashimoto, J.~Yamagishi, and I.~Echizen, ``Privacy-preserving sound to
  degrade automatic speaker verification performance,'' in \emph{2016 IEEE
  International Conference on Acoustics, Speech and Signal Processing
  (ICASSP)}.\hskip 1em plus 0.5em minus 0.4em\relax IEEE, 2016, pp. 5500--5504.

\bibitem{jin2009speaker}
Q.~Jin, A.~R. Toth, T.~Schultz, and A.~W. Black, ``Speaker de-identification
  via voice transformation,'' in \emph{2009 IEEE Workshop on Automatic Speech
  Recognition \& Understanding}.\hskip 1em plus 0.5em minus 0.4em\relax IEEE,
  2009, pp. 529--533.

\bibitem{Bahmaninezhad2018}
\BIBentryALTinterwordspacing
F.~Bahmaninezhad, C.~Zhang, and J.~Hansen, ``Convolutional neural network based
  speaker de-identification,'' in \emph{Proc. Odyssey 2018 The Speaker and
  Language Recognition Workshop}, 2018, pp. 255--260. [Online]. Available:
  \url{http://dx.doi.org/10.21437/Odyssey.2018-36}
\BIBentrySTDinterwordspacing

\bibitem{magarinos2017reversible}
C.~Magari{\~n}os, P.~Lopez-Otero, L.~Docio-Fernandez, E.~Rodriguez-Banga,
  D.~Erro, and C.~Garcia-Mateo, ``Reversible speaker de-identification using
  pre-trained transformation functions,'' \emph{Computer Speech \& Language},
  vol.~46, pp. 36--52, 2017.

\bibitem{pobar2014online}
M.~Pobar and I.~Ip{\v{s}}i{\'c}, ``Online speaker de-identification using voice
  transformation,'' in \emph{2014 37th International convention on information
  and communication technology, electronics and microelectronics
  (mipro)}.\hskip 1em plus 0.5em minus 0.4em\relax IEEE, 2014, pp. 1264--1267.

\bibitem{justin2015speaker}
T.~Justin, V.~{\v{S}}truc, S.~Dobri{\v{s}}ek, B.~Vesnicer, I.~Ip{\v{s}}i{\'c},
  and F.~Miheli{\v{c}}, ``Speaker de-identification using diphone recognition
  and speech synthesis,'' in \emph{2015 11th IEEE International Conference and
  Workshops on Automatic Face and Gesture Recognition (FG)}, vol.~4.\hskip 1em
  plus 0.5em minus 0.4em\relax IEEE, 2015, pp. 1--7.

\bibitem{alegre2014evasion}
F.~Alegre, G.~Soldi, and N.~Evans, ``Evasion and obfuscation in automatic
  speaker verification,'' in \emph{2014 IEEE International Conference on
  Acoustics, Speech and Signal Processing (ICASSP)}.\hskip 1em plus 0.5em minus
  0.4em\relax IEEE, 2014, pp. 749--753.

\bibitem{kaldi_timit}
{Bagher BabaAli and Karel Vesely}, ``{Kaldi TIMIT recipe},''
  \url{https://github.com/kaldi-asr/kaldi/blob/master/egs/timit}.

\bibitem{juvelaniibc}
L.~Juvela, X.~Wang, S.~Takaki, S.~Kim, M.~Airaksinen, and J.~Yamagishi, ``The
  {NII} speech synthesis entry for {Blizzard Challenge} 2016,'' in \emph{Proc.
  Blizzard Challenge Workshop}, 2016.

\bibitem{gussenhoven2004phonology}
C.~Gussenhoven, \emph{The phonology of tone and intonation}.\hskip 1em plus
  0.5em minus 0.4em\relax Cambridge University Press, 2004.

\bibitem{wang2018neural}
X.~Wang, S.~Takaki, and J.~Yamagishi, ``Neural source-filter-based waveform
  model for statistical parametric speech synthesis,'' in \emph{Proc. ICASSP},
  2019.

\bibitem{rec2004p}
I.~Rec, ``P. 563: Single-ended method for objective speech quality assessment
  in narrow-band telephony applications,'' \emph{International
  Telecommunication Union, Geneva}, pp. 1--25, 2004.

\bibitem{Hannun2014DeepSS}
A.~Y. Hannun, C.~Case, J.~Casper, B.~Catanzaro, G.~Diamos, E.~Elsen,
  R.~Prenger, S.~Satheesh, S.~Sengupta, A.~Coates, and A.~Y. Ng, ``Deep speech:
  Scaling up end-to-end speech recognition,'' \emph{CoRR}, vol. abs/1412.5567,
  2014.

\bibitem{nagrani2017voxceleb}
A.~Nagrani, J.~S. Chung, and A.~Zisserman, ``Voxceleb: a large-scale speaker
  identification dataset,'' in \emph{Interspeech}.\hskip 1em plus 0.5em minus
  0.4em\relax ISCA, 2017.

\bibitem{garofolo1993timit}
J.~S. Garofolo, ``Timit acoustic phonetic continuous speech corpus,''
  \emph{Linguistic Data Consortium}, 1993.

\bibitem{prince2007probabilistic}
S.~J. Prince and J.~H. Elder, ``Probabilistic linear discriminant analysis for
  inferences about identity,'' in \emph{2007 IEEE 11th International Conference
  on Computer Vision}.\hskip 1em plus 0.5em minus 0.4em\relax IEEE, 2007, pp.
  1--8.

\end{thebibliography}

\end{document}